# Zoonoses Frontier: Veterinarian, Producer, Processor and Beyond

**Min Yue [1, 2, *], Huanchun Chen[3, 4]**

**[1]The Institute for Translational Medicine and Therapeutics; [2]Department of Pathobiology School of Veterinary Medicine, University of Pennsylvania, Philadelphia, PA 19104, United States of America; [3]Division of Animal Infectious Disease, State Key Laboratory of Agricultural Microbiology, Huazhong Agricultural University, Wuhan 430070, People's Republic of China; [4]Division of Animal Infectious Disease, College of Veterinary Medicine, Huazhong Agricultural University, Wuhan 430070, People's Republic of China**

**To whom correspondence should be addressed: yuemin@vet.upenn.edu**

**Abstract:**

As many emerging and re-emerging infectious diseases are associated with food animals, the relationship between available healthy food sources and population health and social stability has become evident. A recent example of the importance of this relationship was observed during the current flu pandemic. This recent pandemic brought attention to novel target groups of susceptible people at the interface of the animal and human populations. Veterinarians, producers and processors are uniquely exposed to emerging zoonoses. Therefore these individuals may serve as key sentinels and allow efficient evaluation of the effectiveness of zoonoses prophylaxis and control, including evaluation of the cost-effectiveness in the broader view. We also suggest some valuable approaches for rapid diagnosis of emerging and re-emerging infectious diseases and supportive systemic research which may address related ethical questions. We also highly recommend more research investigations characterizing this human/animal zoonosis interface, a potentially productive target for emerging disease diagnosis and control.

**SIGNIFICANCE:**

**In the global environment of emerging and re-emerging infectious disease, zoonoses are a potential threat to human health and economy. The current pandemic influenza outbreak was a significant public health threat that also had great impact on pork production and distribution. The workers at the interface between the animal and human populations provided evidence for their roles in establishment of disease and subsequent transmission. This interface could be a novel sentinel interface for zoonotic disease surveillance and control. Scientific funding and government policy should focus on this zoonotic interface, which could provide a novel approach to control emerging zoonotic disease.**

## Introduction

Infectious diseases, as one of leading causes of death (>25% annual total death worldwide), pose serious threats to population health, animal well being as well as food safety, economic impact and social stability (Morens *et al.*, 2004). Economic development, public health, environmental quality and habitation quality have contributed to the prevention of infectious diseases. Even though such factors as the overall morbidity and mortality have declined slightly, there are still very serious challenges to human health that require global solutions for infectious disease prevention and control. The past few years have seen national and international disease outbreaks of significant veterinary and often zoonotic importance. This is probably the result of several trends, including exponential growth in human and livestock populations, dramatic changes in farming practices, intimate interactions between livestock and wildlife, changes in ecosystems and climate, and globalization of trade of animals and animal products (Tomley & Shirley, 2009, Chomel *et al.*, 2007, Williams *et al.*, 2002). As diagnostic technologies have steadily increased the ability to detect pathogens, more than 1,600 human pathogens have been defined and an average of three new infectious diseases is reported approximately every 2 years. A new infectious agent is described in the medical literature almost every week (Tomley & Shirley, 2009). It has been estimated that the majority of more than 1,400 recognized human diseases are zoonotic and that more than 70 percent of 177 emerging or reemerging diseases have originated from animals (Baker & Gray, 2009, Cleaveland *et al.*, 2001, Tomley & Shirley, 2009). Zoonosis seems the major integrating theme of human and veterinary medicine. This theme of emerging and re-emerging zoonotic infectious diseases has linked world as a “global village”. This linkage indicates that no country can be free of risk in the face of an outbreak of an infectious agent anywhere on the planet. Since the first documented case of the outbreak H1N1 influenza pandemic infection in a Mexico City resident in 2009, the disease has spread to at least 208 countries with 9,596 confirmed deaths as reported by the World Health Organization (WHO) (http://www.who.int/csr/disease/swineflu/updates/en/index.html). Initial genetic characterization suggested swine as the origin of the novel H1N1 influenza virus, on the basis of sequence similarity to previously reported swine influenza isolates (Smith *et al.*, 2009, Garten *et al.*, 2009). However, the generation of this novel influenza involves multiple species and reflects the complexity of the human and animal interface.

## Lessons from Influenza

The recent H1N1 pandemic influenza virus has raised many questions about how animal-origin viruses spread to human populations and evolve as emerging pathogens. From the perspective of infectious disease generation, there are three basic components of this process. Undoubtedly, close human contact with animals provides more opportunity for infection and host transfer. However, direct contact does not always result in infection. Virus persistence in the environment and host innate resistance factors also determine successful disease transmission (Boon *et al.*, 2009, Graham *et al.*, 2008). Veterinarians, abattoir workers, farmers and others closely associated with animal husbandry, with frequent and intense exposures to pigs and poultry for example, are likely to be at elevated risk of zoonotic influenza infection. Previous data indicated that veterinarians and farmers are at risk of infection with influenza virus because of occupational exposure (Kayali *et al.*, 2009, Gray & Kayali, 2009, Gray *et al.*, 2008, Gray *et al.*, 2007). Cross-species transfer infection seems an even more likely event with exposure to large numbers of concentrated animals. In most developing countries, most of the rural population of subsistence producers is involved in small-scale husbandry where livestock, poultry and even pets are housed together. However, farmers practicing small scale or free-ranging poultry and livestock production methods are at an increased risk of infection with influenza virus with a novel tropism and pathogenicity, though there are limited case reports (Graham et al., 2008). Because influenza virus contains eight RNA genomic segments, mixed infections with multiple influenza strains within a production unit may result in genetic reassortment and driving the evolution of novel influenza viruses with the potential to create emerging pandemics. While there is no direct evidence that veterinarians or swine workers played a vital role in the current novel H1N1 pandemic, inevitably, veterinarians and swine workers could potentially serve as a "bridging population" spreading pathogens to their colleagues, families, community and to those animals for which they provide care (Gray & Kayali, 2009). Such events may occur on swine operations because pigs are naturally susceptible to infection with novel type A influenza viruses. What is more, persons who work with swine could play an important role in the mixing of influenza virus strains, directing the adaptive evolution to human beings, leading to reassortment and development of novel progeny strains with pandemic and zoonotic potential (Gray & Kayali, 2009). The primary swine or poultry care givers are potentially among the first to be infected in the event of an emerging virus becoming epizootic among swine herds, and those who work with swine may serve as a bridge for transmission of the virus to their communities. Persons who work with swine could also be considered for sentinel influenza surveillance and the early diagnosis of infection could indicate a potentially emerging endemic or pandemic pathogen within the population (Gray & Kayali, 2009). However, current policies to prevent an influenza pandemic often overlook veterinarians and animal workers. Surveillance of these individuals is traditionally neglected and pathogen biology and ecology should focus on the workers at the animal-human interface. Policy measures and biosecurity procedures should be implemented to reduce hazards for veterinarians and animal workers that could prevent transmission of zoonotic diseases to others human and animal populations.

**Multiple Roles of "Bridging Groups"**

"Bridging groups" are people whom are in close contact with or have direct exposure to animals and animal products. These groups include veterinarians, farmers, and abattoir workers. A broader definition would include those individuals that could link an emerging disease from animals to the human community such as field workers, outdoor and wildlife enthusiasts, zoo keepers and pet owners. Their relationships with zoonotic disease are described in Figure 1. For influenza virus, veterinarians are just one important segment of the zoonotic disease frontier. The workers living and/or working on large scale progressive production units, people in developing countries in close proximity to their various domestic animals (usually pigs, chickens ducks, geese and water buffaloes), workers at processing facilities should all be considered as potential "bridging groups" for transmission and spread of influenza as well as other zoonotic diseases (Sahani *et al.*, 2001). Those who have frequent exposure to wildlife are also a potential surveillance target group for emerging zoonotic diseases (Moll van Charante *et al.*, 1998). Also, pet owners and zoo workers could be included as potential bridging groups. People who are chronically ill and possibly immunosuppressed but dependent on animals for companionship and/or for their livelihoods may also represent a unique set of bridging groups (Chomel et al., 2007, Trevejo *et al.*, 2005).

In the past research focused on either the animal or human population for zoonotic disease research, particularly with regard to diagnostics, pathogenesis and clinical prevention or management. While these populations are individually very important for disease surveillance and control the people in direct contact with the animals may better serve as sentinels for emerging infectious diseases. This concept provides an opportunity for reconsideration of the zoonotic disease frontier between humans and animals. The appreciation of pathogen ecology at the human/animal interface may facilitate recognition of emerging and reemerging disease events. Previous reports include numerous publications about occupational disease among workers n animal agriculture that focus on the chronic diseases associated with environmental factors and host atopy. There is little information regarding relative risk of contracting zoonotic disease, whether in reference to an established disease or an emerging, novel infectious disease (Hoppin *et al.*, 2003, Radon *et al.*, 2001). Sometimes individuals in the bridging groups may have specific immunity to pathogens and exhibit few clinical signs and may become a healthy carrier and transmitter of infection. This asymptomatic colonization and shedding could increase the potential of the carrier individual as a threat to the general public health. Therefore, surveillance of these bridging groups seems to be a promising frontier in human health and this deserves considerable attention. Because of the unpredictability of pathogen emergence, the first line of defence has to be aggressive and effective surveillance, requiring identification and monitoring of high-risk populations or individuals - the “bridging populations” seem to be the ideal target for disease identification and monitoring.

## Additional Suggestions for Related Research

Based on these considerations, the following two major objectives have been identified as areas of possible emphasis by national authorities and scientists who are engaged in infectious disease research and surveillance. The first goal is the enhancement of available resources for disease surveillance and subsequent responses to emergencies. For known infectious agents, we need to focus on the specific agents as targets for molecular and serological surveillance within the “bridging population”. Monitoring and surveillance programs should be developed according to conditions present in individual countries or regions. For very rare and/or low risk disease situations, passive monitoring and surveillance approach of reporting of clinical suspect and confirmed cases is likely sufficient. For more common and higher risk diseases, active monitoring and surveillance along with regular, periodic collection of case reports can be applied to epidemic area (Doherr & Audige, 2001). Novel rapid, integrated assays for specific dangerous infectious agents with high throughout capability based on population requirements are urgently need. Common primers for PCR and sequencing may represent an ideal way for pathogen screening and identification as well as for epidemiological investigations. Microarray and metagenomics may also be applied to simultaneously detect multiple pathogens of interest. Epidemic early warning mechanisms and preparedness plans are also needed to complete the surveillance programs. More importantly, national authorities play a key role in devising, financing and implementing these intervention tools.

The second goal is basic research in pathogen biology and the continued discovery and characterization of new infectious agents. A better understanding of the factors controlling the emergence and spread of infectious diseases is needed. The "jumping" mechanisms for cross-species infection and spreading need to be understood. Strategic research to enable targeted disease control programs which can be applied to disease control is essential. The exploration of previously unknown agents, including those agents with new host tropisms, biological phenotypes or causing new clinical presentations can provide valuable information for understanding disease transmission, pathogenesis and control. Projects such as the current Human Microbiome Project can provide tremendous information for understanding emerging diseases. These data can also be applied to disease surveillance and potential risk evaluation within target animal populations. For the "bridging group", the metagenomic approach may be an ideal approach in the HMP for surveillance and potential risk evaluation.

**Questions remain and Future direction**

The concept of "bridging groups" represents a novel concept from Dr. Gray (Gray, et al 2008). The first and greatest obstacle in utilizing these potential surveillance populations is the existing scientific limitations. Though we have evidence about zoonotic disease transmission directly from animals to humans, there are also other unknown factors that need to be understood. The second obstacle is evident in the decision-making process of disease surveillance. Though medicine has classic concepts of infectious disease control, these concepts must be considered from the perspective of the “bridging group”. The cost-effectiveness of the control programs must also be evaluated within the “bridging group”. The range and numbers of people included for regular monitoring, possible vaccination and other intervention measures in specific disease control programs must also be evaluated (Gray & Kayali, 2009, Murphy, 2008). The addition expenditures and effort associated with total disease control and surveillance may impact the human rights and ethics of the specific occupational population association with animal agriculture. Therefore this is more than a scientific problem. Surveillance of the “bridging groups” may also have direct impact on safety of animal products, which is another important route for zoonotic disease transmission.

It is possible to have an even higher order of baseline preparedness through animal surveillance of a range of pathogens before they have the opportunity to infect and spread disease to humans. Initiating preventive actions by dealing with the causes and drivers of infectious diseases, particularly at the animal–human–ecosystems interface, seems also to be a promising approach. With regard to zoonosis control and prevention, integration and analysis of the surveillance data from both animal and human populations with simultaneous analysis would facilitate early warning and risk reduction. Systematic pathogen surveillance of animals and recognition of changes in pathogen prevalence, host range tropisms and pathogenesis or virulence as signals at the herd or production unit level could presage an emerging disease and possibly minimize the consequences to human and animal health. Neglected groups which are in close contact with animals due to occupational exposure need considerable more surveillance and disease prevention efforts to block an emerging zoonotic disease situation (Myers *et al.*, 2007). These approaches seem a better choice for zoonotic disease prevention. It is highly unlikely that there will be any way to predict when or where the next important, new zoonotic pathogen will emerge; nor will there likely be any way to predict a new pathogen's ultimate importance from its early behavior. The health and safety of the animal and human populace depends on the continuous ability to rapidly detect, monitor and control newly emerging or re-emerging livestock disease and zoonoses rapidly (Doherr & Audige, 2001).

Zoonotic diseases require multidisciplinary and comprehensive research studies. Multi-sectoral collaboration and policy-oriented discussion will allow disease prevention and control. Where specific occupational exposure and close contact with animals or animal products, this discussion provides some information about how to do the supportive research and subsequent disease prevention and control on zoonotic diseases. Specific strengthening of research and testing efforts on the "bridging populations" promises a new and hopeful frontier for early warning and control of emerging zoonotic diseases. Different countries may have different levels of surveillance and control of specific diseases. Nevertheless, collaboration and cooperation will allow new surveillance and control programs to prevent the emergence of pandemic disease from the bridging populations at the interface of animals and humans.

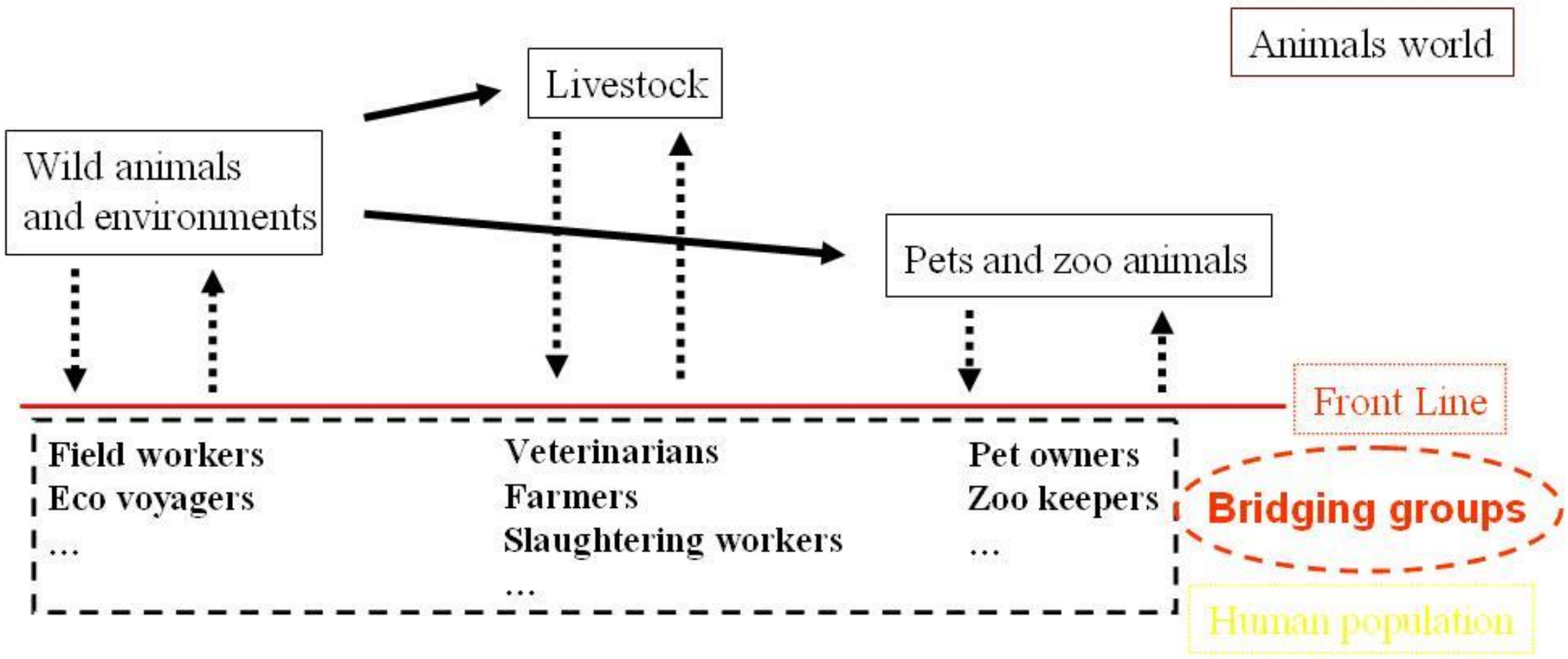

**Fig.1 Overview of the zoonoses front line and the "Bridging groups".**

Solid arrows indicate the EID or zoonoses pathogens transmission in animals. Dotted arrows indicate the zoonoses pathogens transmission between human and animals. The figure specifically emphasizes the zoonoses transmission interface, which was defined as the "Bridging groups". There are conclusive three major pathways for zoonoses disease contribute to "Bridging groups" at animal-human interface. Environmental pathway through wild animals and environments exposure, occupational pathway though exposure to industrial-scale livestock operation and lifestyle pathway during daytime exposure to companion animals.

## Acknowledgement

This work was supported by the 973 Program (grant no. 2006CB504404) and Innovation Teams of Ministry of Education (grant no. IRT0726). All the authors would like to thank D Scott McVey for his critical comments.